\documentclass[aps,prf,showpacs,floats,twocolumn,floats,superscriptaddress,floatfix,longbibliography]{revtex4-1}
\usepackage{bm}
\usepackage{hyperref} 
\usepackage[usenames,dvipsnames]{xcolor}
\usepackage[normalem]{ulem}
\usepackage[utf8]{inputenc}
\usepackage{amsmath}
\usepackage{amsthm}
\usepackage{amsmath}
\usepackage{amsfonts}
\usepackage{amssymb}
\usepackage{blindtext}
\usepackage[utf8]{inputenc}
\usepackage[T1]{fontenc}
\usepackage{graphicx}
\usepackage{float}
\usepackage{wrapfig}
\usepackage{bm}
\usepackage{braket}
\usepackage{grffile}
\usepackage[normalem]{ulem}
\usepackage[usenames,dvipsnames]{xcolor}
\usepackage{soul}

\def\RMV#1{{}}

\newcommand{\be}{\begin{equation}}
\newcommand{\ee}{\end{equation}}

\newcommand{\Xtargett}{{\bold X}_t}
\newcommand{\Xagentt}{{\bold X}^{(a)}_t}

\newcommand{\bphi}{\bm{\phi}}
\newcommand{\bn}{{\boldsymbol n}}

\newcommand{\bx}{\boldsymbol{x}}

\newcommand{\bu}{{\boldsymbol u}} 
\newcommand{\bU}{{\boldsymbol U}}

\begin{document}
\title{Optimal tracking strategies in a turbulent flow 
}

\author{C. Calascibetta} \email{calascibetta@roma2.infn.it}
\affiliation{Department of Physics \& INFN, University of Rome ``Tor
Vergata'', Via della Ricerca Scientifica 1, 00133 Rome, Italy.} 
\author{L. Biferale}
\affiliation{Department of Physics \& INFN, University of Rome ``Tor
Vergata'', Via della Ricerca Scientifica 1, 00133 Rome, Italy.}
\author{F. Borra}
\affiliation{Laboratory of Physics of the École Normale Supérieure, 24 Rue Lhomond, Paris, 75005, France.}
\author{A. Celani}
\affiliation{Quantitative Life Sciences, The Abdus Salam International Centre for Theoretical Physics, ICTP, Trieste, 34151, Italy.}
\author{M. Cencini}
\affiliation{Istituto dei Sistemi Complessi, CNR, Via dei Taurini 19, Rome, 00185, Italy, and INFN `Tor Vergata'.}

\date{\today}

\begin{abstract}
 Pursuing a drifting target in a turbulent flow is an extremely difficult task whenever the searcher has limited propulsion and maneuvering capabilities. Even in the case when the
  relative distance between pursuer and target stays below the turbulent dissipative scale, the chaotic nature of the trajectory of the target represents a formidable challenge. Here, we show how to successfully apply optimal control theory to find navigation strategies that overcome chaotic dispersion and allow the searcher to reach the target in a minimal time. We contrast the results of optimal control -- which requires perfect observability and full knowledge of the dynamics of the environment -- with heuristic algorithms that are reactive -- relying on local, instantaneous information about the flow. While the latter display significantly worse performances, optimally controlled pursuers can track the target for times much longer than the typical inverse Lyapunov exponent and are considerably more robust.
\end {abstract}

\maketitle
 Finding optimal navigation strategies in a complex fluid environment is a notoriously difficult problem with applications ranging from environmental monitoring \cite{trincavelli2008towards,zhang2008optimal,Bellemare2020,monitoringoceanbiogeochemistry} to
micro-medicine \cite{wang2012nano,li2017micro,wang2021trends}.
A well-explored set of navigation problems is point-to-point path-planning optimization of flying vehicles such as
airplanes or drones, with the aim of  minimizing some functioning cost that may comprise fuel consumption and time of arrival \cite{szczerba2000robust,SONG2017388,Guerrero2013}. These vehicles
move in a complex chaotic environment but can have almost full control on their trajectory as their speed is typically  larger than the fluid velocity. Recently, point-to-point path-planning
optimization has been the focus of intense research also for slow, microscopic objects, such as microswimmers or active particles, which tend to be carried away by the flow and need to appropriately exploit it in order to reach their destination
\cite{Nasiri_2023,Lolla2014,RHOADS201312,biferale2019zermelo,Buzzicotti_Zermelo2020,alageshan2020machine,daddi2021hydrodynamics,gunnarson2021learning,Verma_2018,Goh_2022,Zhu_2022,yang2018optimal,yang2020efficient,piro2022optimal,piro2022efficiency,CalascibettaEPJE,xu2023long}.

\begin{figure*}[t]
\includegraphics[scale=.35]{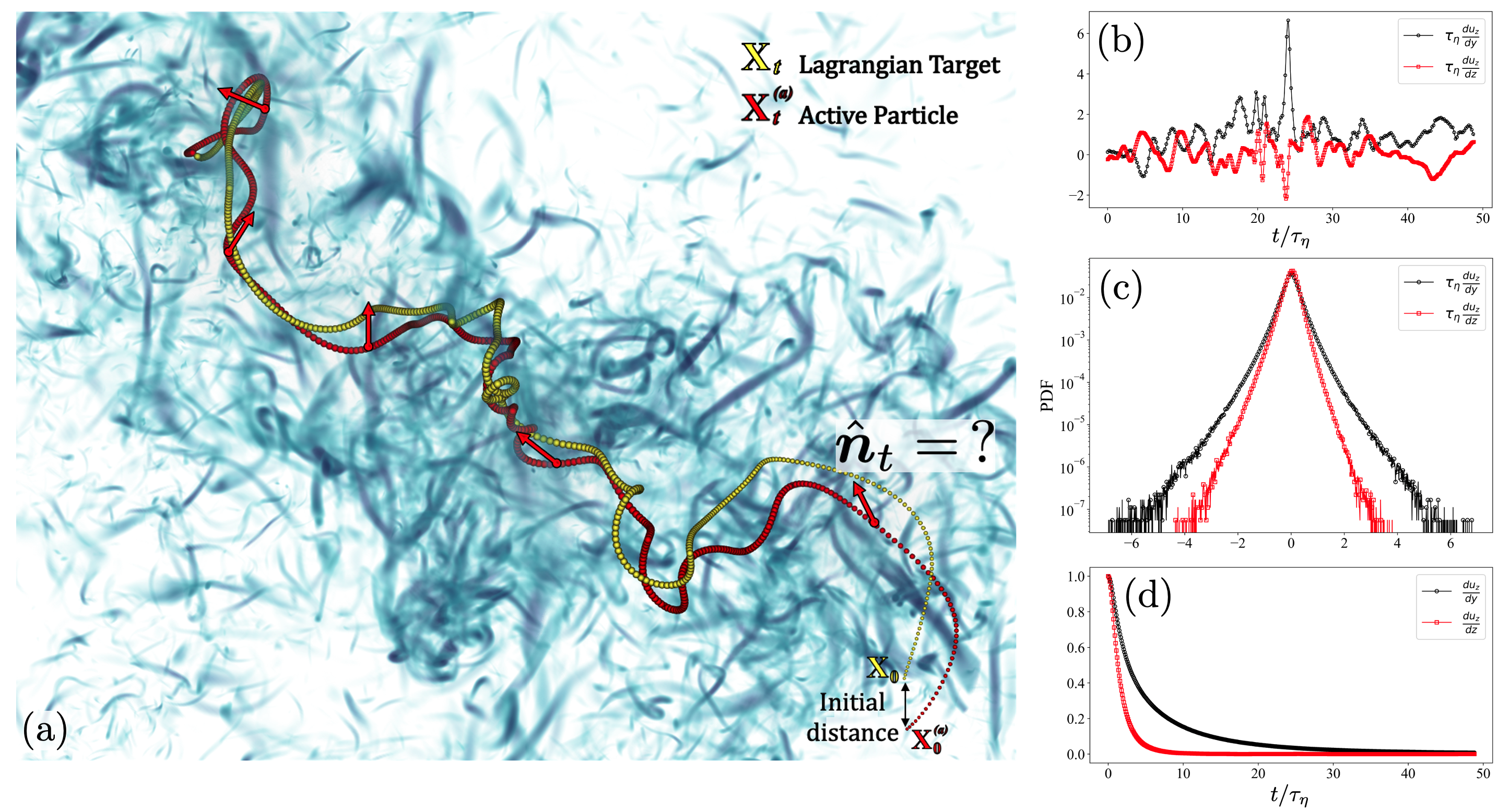}
\caption{  \textbf{(a)} Artistic summary of the problem. A pursuing agent (in red color) with limited maneuverability aims to stay as close as possible or, possibly, to capture a Lagrangian target (in yellow color) chaotically advected by a turbulent flow. The background is given by a rendering of the turbulent vorticity intensity at a fixed time during the episode. \textbf{(b)} Time evolution of the typical dimensionless transverse, $du_z/dy$,  and longitudinal, $du_z/dz$,  velocity gradients during the duration of a catching episode, where the normalization factor is given by the Kolmogorov time-scale of the flow, $\tau_\eta =\sqrt{\nu/\epsilon}$, where $\nu$ is the fluid viscosity and $\epsilon$ the mean energy dissipation. \textbf{(c)} Probability distribution function (PDF) of the transverse and longitudinal gradients measured over  $200$K trajectories of length ${\cal T} \simeq 150\tau_\eta$. \textbf{(d)} Time correlation functions for the transverse and longitudinal velocity gradients.  \label{fig:artistic}  } \end{figure*}

Here, we consider a more difficult navigation task, where the target is not fixed in space but is chaotically advected by the turbulent flow. The challenge of tracking a Lagrangian target is increased by the limited speed and manoeuvrability of the pursuer. 
This problem is relevant to many applications, such
as keeping in a pattern formation a swarm of oceanic
drifters and floaters \cite{Peterson2011,Song2015,SONG2017388,Mallory_2013,WYNN2014451}; 
interpreting strategies to catch non-swimming
preys by micro-swimmers in turbulent environment; developing
autonomous self-propelling protocols for mini-robots navigating in
complex bio-flows or for gliders in the atmosphere or oceans \cite{Bellemare2020,dunbabin2008,smith2011,lumpkin_pazos_2007,RevModPhys.88.045006,PhysRevLett.121.078001,Popescu_2011,C1SM06512B,IJAC-2019-05-105}.
We will consider the case in which searcher and target stay at distances smaller than the Kolmogorov length and thus experience a smooth chaotic flow at all times. As depicted in Fig.~\ref{fig:artistic}a, due to its limited speed the agent must know how to {\it surf} local eddies in ingenious ways, by taking advantage of strong fluctuations, sometimes generated by vortical structures (panels b-c), and exploiting the long time correlations typical of turbulence (panel d).   The final objective of the optimal pursuer is to catch the chaotic moving target in the shortest possible time or, if all else fails, to be as close as possible to it at the end of the allotted time for the pursuit.\newline\newline
\begin{figure*}[t]
\includegraphics[scale=0.34]{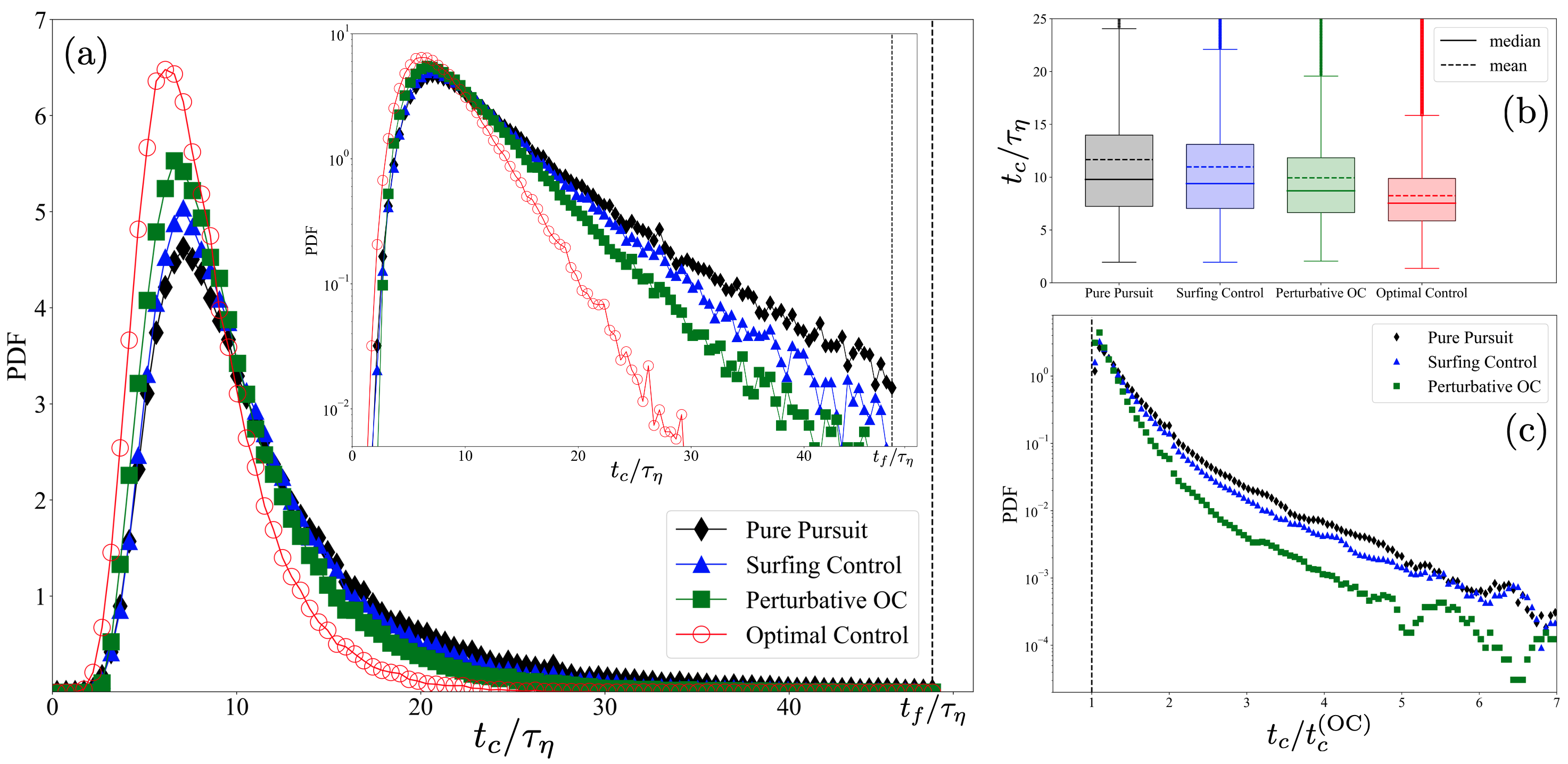} \caption{\textbf{Capture time statistics.} \textbf{(a)} PDF of the capture time for the Optimal Control strategy (red open circles) and for the heuristic strategies, Pure Pursuit (black full rhombus), Surfing Control (blue full triangles) and Perturbative Optimal control (green full squares). The PDFs are evaluated along the episodes where all strategies capture. The vertical dashed line represent the time horizon, $t_f$. In the inset we show the same PDF in log-scale. \textbf{(b)} 
Box-plot of the catching time PDF. The box shaded area reports the range 25th-75th percentile, the solid and dashed lines are the median and the mean respectively. Each whisker contains the remaining $25\%$ of the data, while outliers points are identified for large value of the capture time.   \textbf{(c)} PDF of the  capture time for the reactive heuristic strategies, normalized with the corresponding capture time for the OC, $t_c/t_c^{(\text{OC})}$ constrained to the episodes where all the heuristic strategies capture within $t_f$.}
\label{fig:2}   
\end{figure*}
The main obstacle in finding the optimal control is that it depends on the entire spatio-temporal story of the chaotic system evolution. Analytical solutions can be found only in simple linear or time invariant flows, limited to the point-to-point navigation task \cite{bryson,OC_Aerospace,daddi2021hydrodynamics,liebchen2019optimal,Hays2013}.
In this paper, we show how to apply Optimal Control Theory to discover the best controls for the turbulent tracking problem and  
compare them with heuristic reactive strategies.
In particular, since the problem is formally akin to a
pursuit game we use for comparison heuristics like the (i) pure pursuit
\cite{nahin2012chases}, i.e. always swimming along the
line-of-sight of the target, and (ii) two other
heuristic controls obtained by  solving simplified 
optimality problems over a short time horizon with the help of some assumptions on the dynamics.  \color{black} We show that
such heuristic strategies, which certainly have the advantage of not requiring a large amount of information on the system, are markedly sub-optimal. How to develop effective heuristic strategies with limited information and computation but whose performances are closer to optimality remains an outstanding problem.

\section{Results}
\textbf{Tracking Lagrangian targets.} 
The pursuer -- hereafter also called ``the agent" -- and the target are immersed in a turbulent velocity
field $\bu_t(\bx)$ obtained by direct numerical integration of the
Navier-Stokes equation, sustained by an isotropic and homogeneous
forcing, with Reynolds number at the Taylor scale $Re_\lambda \simeq 130$ \cite{frisch1995turbulence,pope2000turbulent} (see ``Methods'' for
details).  Episodes start with the target, $\Xtargett$, and the agent, $\Xagentt$,  placed at a distance of the order of the Kolmogorov scale, $|{\mathbf X}_{0}^{(a)}\!-\!{\mathbf X}_{0}|\!=\!|\bm R_{0}|\!=\!R_{0}\simeq \eta$.  The flow transports both the target, which
moves as a passive tracer, and the agent that, however, can exert some
control by swimming, with velocity $\bU_t$, with respect to the medium. While it can freely point in any
direction $\hat{\bn}_t$, the ``swimming velocity" is constrained to have a fixed speed, $V_s$. We chose this speed
to be smaller than the Kolmogorov velocity, $u_\eta$. Specifically, the results shown below are for $V_s \simeq 0.13 u_\eta$.

The two point-like objects therefore move according to the dynamics
\begin{equation}
    \begin{cases}\label{eq:dynamics}
\dot{ \bold X}_t =  \bu_t(\Xtargett)  \\
\dot{ \bold X}^{(a)}_t =  \bu_t( \Xagentt) + \bU_t \\
\bU_t = V_s \hat{\bn}_t\,,
\end{cases}
\end{equation}
where  we consider the reorientation time of the control direction negligible.
The episode lasts up to a maximum time $t_f \simeq 50 \tau_\eta$, i.e. much larger then the inverse of the Lagrangian Lyapunov exponents of
the tracer trajectory $\lambda^{-1} \simeq 7.5 \tau_\eta$ \cite{cencini2010chaos}, where
$\tau_\eta$ is the Kolmogorov time. Within this time horizon
the agent has to find the optimal choice of the steering protocol
$\hat{\bn}_t$ that allows to capture the target. The capture event is defined as the moment when the relative distance becomes smaller than the capture distance $R_c
=10^{-2}R_{0}$, which is much smaller than the initial separation. From that time on, the
agent sticks to the target.  The speed is set as
$V_s\approx \lambda R_{0}$ so that the agent is at the limit of
controllability, i.e. the control is  smaller than the typical background velocity between the agent and target.

As long as the agent and target remain within distances where the flow is differentiable, i.e. , $R_t \leq O(10\eta)$, the
fluid velocity experienced by the searcher is approximately $
\bm u_t(\bold X^{(a)}_t)\simeq \bm u_t(\bold X_t)+\nabla \bm u_t
{\bm R}_t$, where $\nabla \bm u_t$ is the flow gradient evaluated at
the target position. As a result, the optimization problem
depends only on the separation vector ${\bm R}_t$ and on the entire
history of the velocity gradients along the trajectory of the
searching agent. We will be enforcing this approximation, which allows
us to store Lagrangian trajectories and their accompanying velocity gradient to create a database that can be used to compute the optimal control (see ``Methods'').

\textbf{Heuristic control strategies.} 
As a term of comparison, we will consider several heuristic controls that neither require knowledge of the future evolution of the trajectory (required for the computation of the optimal control) nor the memory of its past. These are called reactive strategies as they are purely based on instantaneous information about the flow and the pair separation.
\begin{enumerate}
\item \textit{Pure Pursuit (PP).} 
The agent only knows in which direction the target currently is and
swims towards it \cite{nahin2012chases}:
\[\hat {\bm n}^{\text{PP}}_t = - \hat{\bm
  R}_{t}       \,.
  \]
  \item \textit{Surfing Control (SC).} The agent has knowledge of the instantaneous velocity gradient and direction of the target \cite{monthiller2022surfing}: 
\[
\hat {\bn}^{\text{SC}}_t \propto
-[\exp(\tau_s \nabla \bu_{t})]^\intercal \hat{\bm R}_{t}\,,
\]
where $\tau_s$ is a parameter to be chosen empirically.
This control maximizes the displacement of the pursuer towards the target in a time interval $\tau_s$, assuming that the gradient and the direction of the target $\hat{\bm R}_t$ are constant over that time (see Methods).
\item \textit{Perturbative Optimal Control (PO).}    The agent has the same information as in (SC) but the direction in which it swims is   
\[
\hat{\bm n}^{\text{PO}}_t \propto -[\exp({\tau_p{\nabla
      \bu_{t}}})]^\intercal \exp( \tau_p {\nabla \bu_{t}})\hat{\bm R}_t\,.
\]
 As discussed in Methods, we obtained this new control strategy by solving perturbatively
 the optimal control problem which minimizes the distance
from the target after a time $\tau_p$, assuming constant gradients (but dynamically evolving separations). The time $\tau_p$ is a free parameter to be chosen empirically. 
\end{enumerate}

\textbf{Optimal control.} 
How can one define the performance of a tracking agent? Ideally, the searcher should be able to capture the moving target in the shortest possible time, and in any event before a certain time horizon. However, this is not always possible because of the strong dispersion induced by the underlying turbulent flow. In these situations one could instead settle for the less ambitious objective of reaching the shortest possible separation at the end of the allotted time for the search. These requests can be summarized by the following mathematical expression for the cost function
\begin{eqnarray}
\label{eq:oc_J}
J = R^2_{t_f} + c\lambda R_c^2 \int_{0}^{t_f} dt\, \Theta(R_t- R_c)\,.
\end{eqnarray}
The first term of the cost function $J$ is the distance between agent and
target at the final time while the second one, $\Theta$ being the Heaviside function,
is the time needed to reach the capture distance, $R_c$. 
If the episode ends with a capture
for $t<t_f$ the first term equals $R_c^2$ and does not depend on the control so that the task is to minimize the time for capture. Conversely, if the capture does not occur before $t_f$ the second term is always equal to $c\lambda R_c^2 t_f$ regardless of the trajectory and the task reduces to minimizing the final separation $R_{t_f}$. The
constant $c$ controls the trade-off between the two contributions and it is chosen  to give a strong enough weight to the capture events.

The theory of Optimal Control (OC) provides the tools to compute the best control -- i.e. the one with minimal cost -- by casting the optimization in the form of Euler-Lagrange equations that have to be solved both forward and backward in time, requiring full knowledge of the dynamics of the system (see Methods) \cite{bryson,lenhart2007optimal,trelat2012optimal}. If a solution to the extremality conditions exists and is stable, it provides at each time the optimal direction $\hat{\bm n}^{\text{OC}}_t$ which implicitly depends on the whole history -- past, present and future -- of gradients.

\begin{figure}[b!]
\includegraphics[scale=.23]{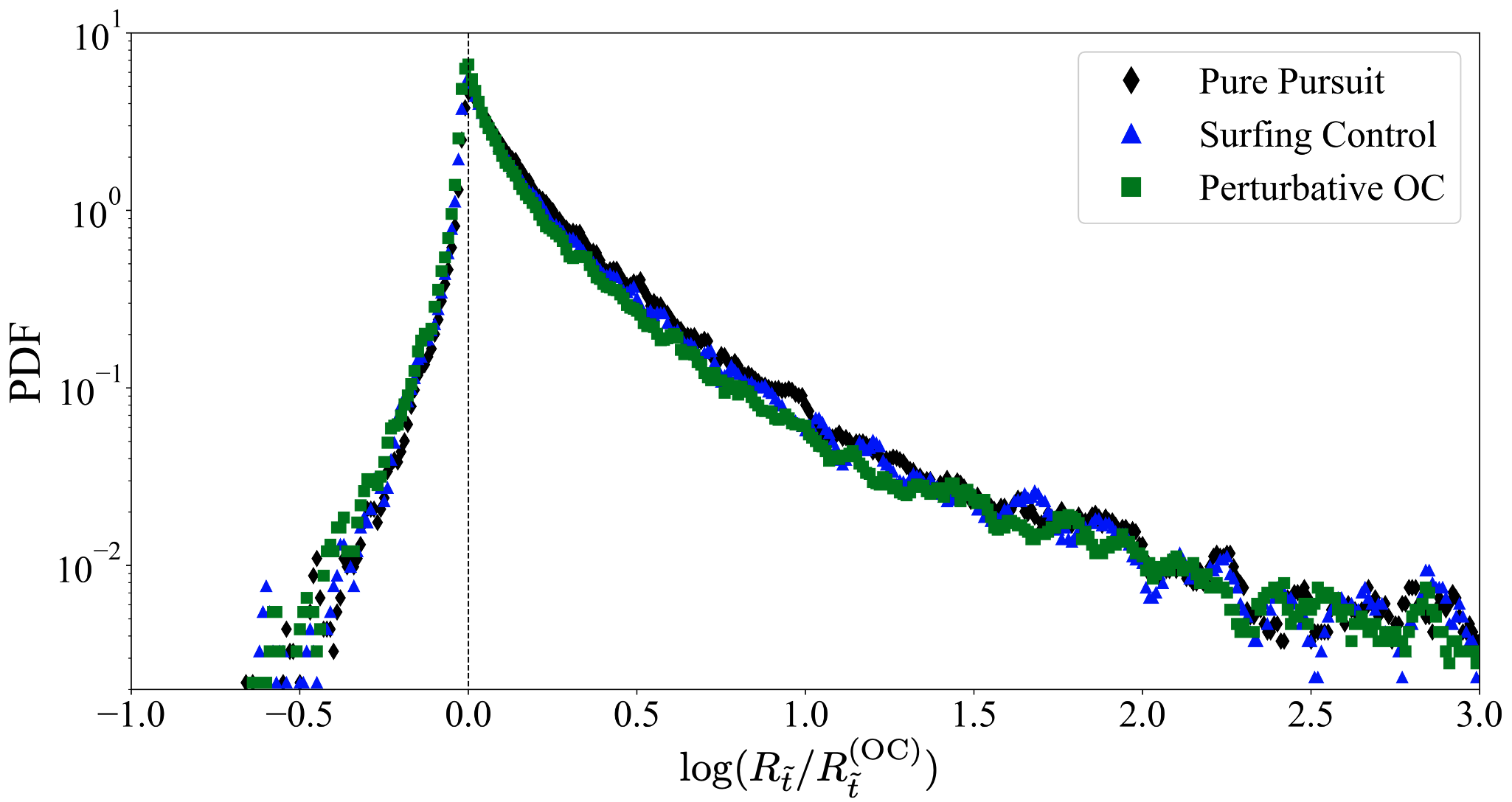}
\caption{\textbf{Statistics of final distances.} PDF of the logarithm of the distances reached by all reactive  heuristic strategies, $R_{\tilde t}/R_{\tilde t}^{(\text{OC})}$  at the time, $\tilde t$, when the OC trajectories trespass the value $10 \eta$. The statistic refers to the episodes where OC fails in catching the target. }
\label{fig:pdf_finaldist}       
\end{figure}

\textbf{Capture time statistics.} We sampled $2\times 10^5$ Lagrangian targets trajectories with the agent starting at a random position distant $R_{0} \simeq \eta $ from the target. The rate of convergence of our OC algorithm  is close to $80\%$. Among the optimally controlled paths, the target is successfully captured in $81\%$ of the cases. With further fine-tuning of the hyper-parameters involved in the iterative process, one might still improve the convergence rate of the algorithm up to probably $100\%$. Over the same converged set, the percentages of success of the three heuristic strategies are similar and close to $70\%$.  A clear supremacy of the OC strategies is found when looking at the PDF of catching times, $t_c$, shown in Fig.~\ref{fig:2}a. Here, we see that OC protocols considerably deplete the probability to search for long times, as shown by comparing the far right tails for $t_c>t_m$ ,with $t_m\simeq 9\tau_\eta$ the mean capture time for the OC. Furthermore, considering the remaining $10\%$ of episodes where OC captures but the heuristics are unsuccessful, we find that OC is also able to capture at much longer timescales, even close to the time horizon (not shown). In  Fig.~\ref{fig:2} we  show that the OC protocol provides a  systematic advantage also when comparing the mean catching time (panel b) and a  strong improvement (up to $\times 6$) for the rarest events when compared trajectory-by-trajectory against each of the heuristic reactive strategies (panel c).
Furthermore, all heuristic reactive strategies perform similarly, stressing the intimate limitations of strategies based on local-in-time cues.

\begin{figure}[t]
\includegraphics[scale=0.27]{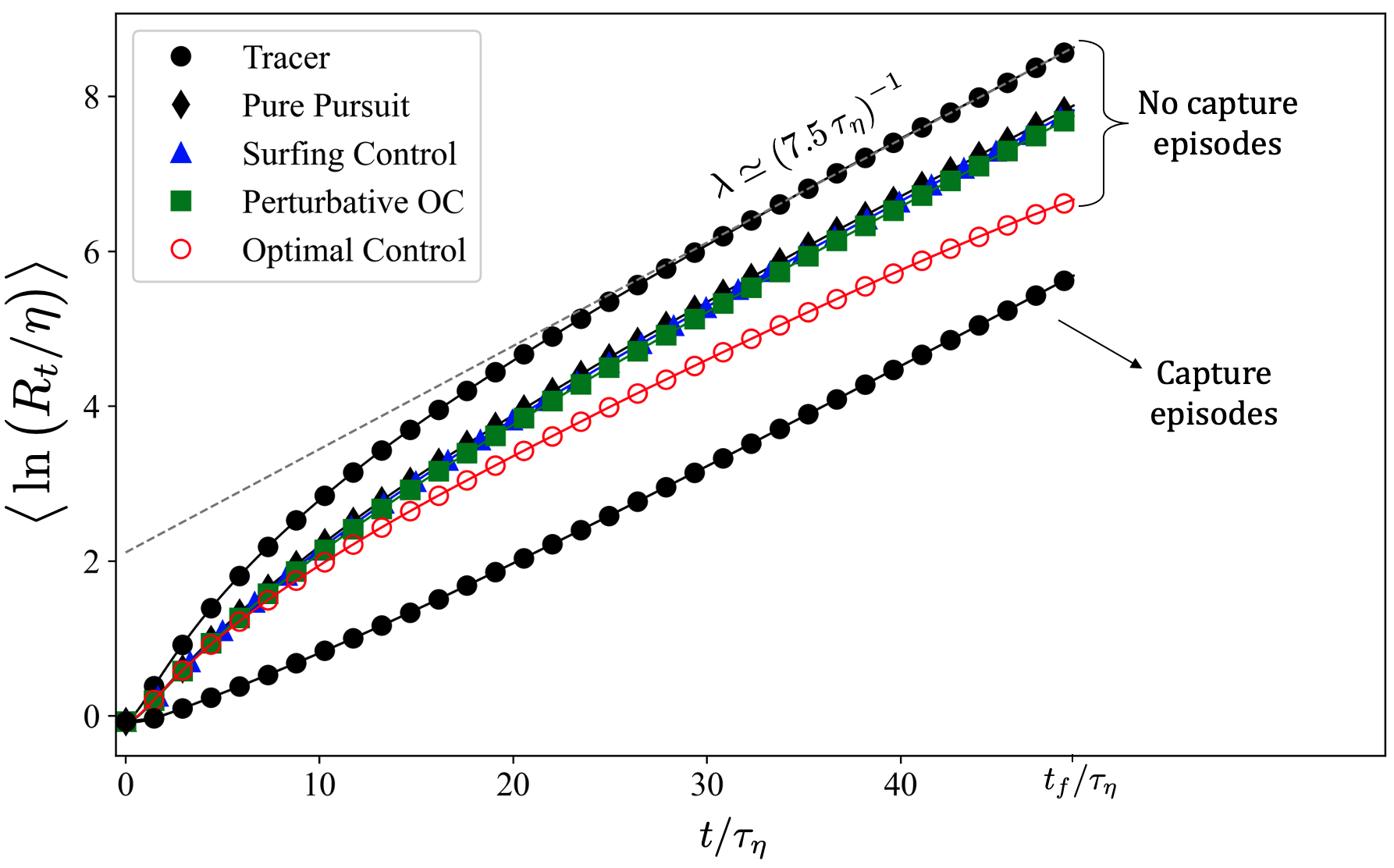}
\caption{\textbf{Logarithm of the error growth averaged over the trajectories for all  strategies.} Pure Pursuit (black full rhombus), Surfing Control (blue full triangles), Perturbative Optimal control (green full squares) and OC (red open circles) are measured only on the unsuccessful episodes (no capture).  The growth rate of the uncontrolled searcher (the tracer with black full circles) is shown for both capture and no-capture episodes.}
\label{fig:separation}       
\end{figure}
\textbf{Final and intermediate distance statistics.} 
When the target cannot be captured within the allowed time horizon OC is still optimal with regard to the final distance from the target, as imposed by the first term in the RHS of (\ref{eq:oc_J}), and almost optimal also at intermediate times. The latter  is shown in Fig.~\ref{fig:pdf_finaldist}, where the PDF of the normalized separation, $R_{t}/R^{(\text{OC})}_{t}$, for each of the heuristic strategies is shown at the time when the corresponding OC trajectory reaches the separation $R^{(\text{OC})}_{\tilde t} = 10\eta$. Except for a small set of events, the OC solution is always closer to the target as shown by the strong asymmetry between left and right tails. 
\begin{figure*}[t]
\includegraphics[scale=0.45]{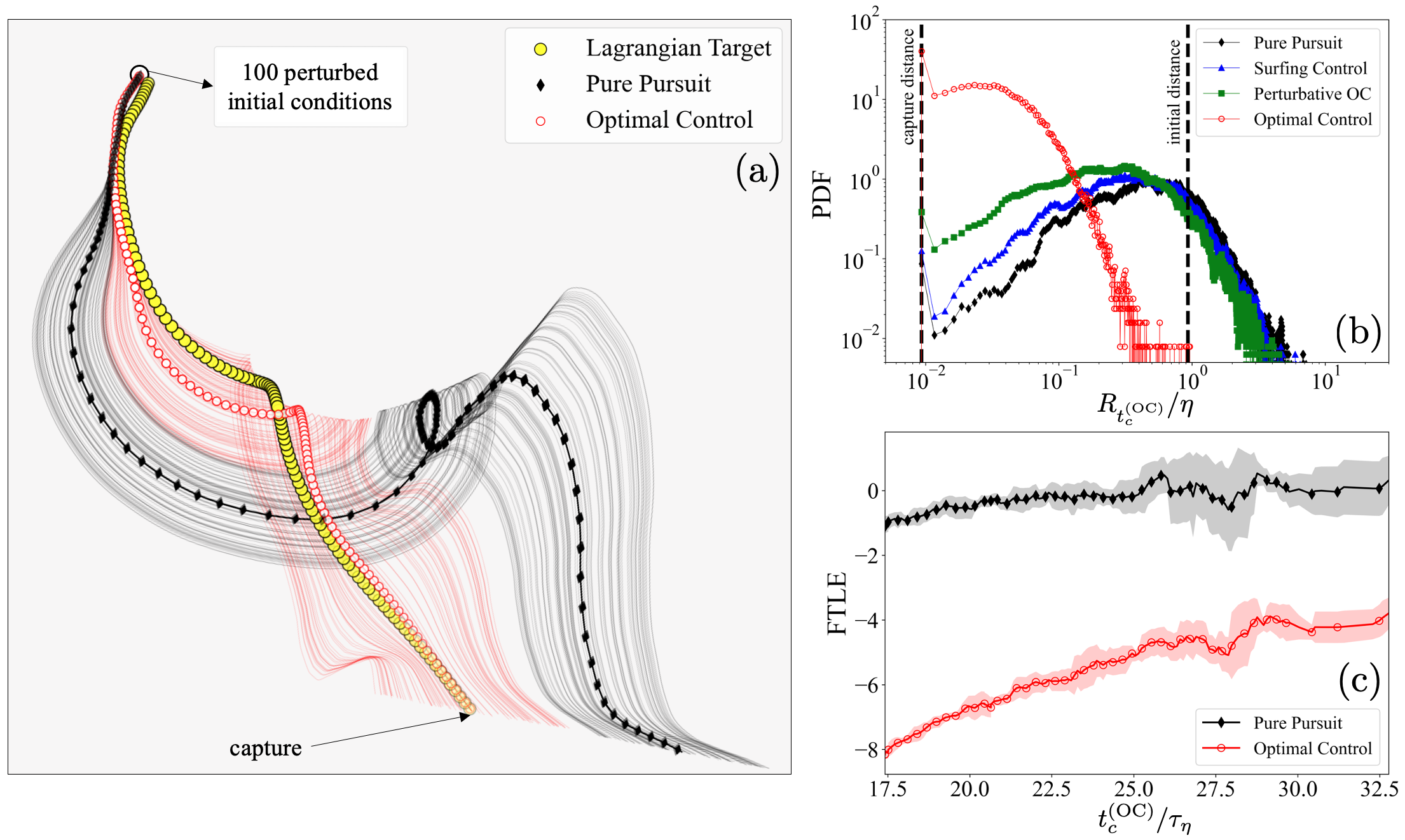}
\caption{\textbf{Robustness under perturbations of the initial conditions.} \textbf{(a)} Example of 100 trajectories obtained by varying the initial conditions of the pursuers. Target reference trajectory (large yellow circles), unperturbed  OC trajectory (red open circles), Pure Pursuit strategy (black full rhombus). Lines show the perturbed trajectories, starting with an error $\delta R_{0} \simeq \eta/10$, same colors of the unperturbed reference trajectories.  \textbf{(b)} PDF of the distances reached at the catching time $t_c^{(\text{OC})}$ by each control strategy starting from perturbed initial conditions. The PDF refers to episodes where OC captures in times larger than the average capture time, i.e., $t_c^{(\text{OC})}\gg t_m^{(\text{OC})}$ and the heuristic strategies capture within the time horizon, $t_f$. The error on the initial position is set to $\delta R_{0}=\eta/100$. The two vertical dashed lines from left to right indicate the capture and the initial distance respectively. \textbf{(c)} FTLE for OC and the Pure Pursuer perturbed strategies evaluated as a function of the unperturbed capture time $t_c^{(\text{OC})}$ for the same episodes of (b). The black full rhombus and the red open circles indicate the mean over all perturbed trajectories, while the shaded areas indicate their standard deviations.}
\label{fig:sensitivity}       
\end{figure*}

\textbf{FTLE and controllability}. To understand the conditions for a successful capture,  it is useful to look at the time-evolution of the typical growth of separations along the trajectories of the target and conditioned on the capacity of the OC protocol to succeed/fail in the  capture (black full circle in Fig.~\ref{fig:separation}). Asymptotically, the growth rate is the same for all protocols and given by the Lagrangian Lyapunov exponent $\lambda$ (dashed straight line).  For short time, $t\lesssim 10\tau_\eta$, we see that the two sets behave quite differently. In particular, the unsuccessful episodes are characterized by a growth rate larger than the average (i.e. $>\lambda$) while
the successful ones remain close to the average and even slightly below at the very early times.  
In the same figure, we also show along  the unsuccessful episodes the growth of the distance  between the target and the (controlled) agent for the heuristic and OC protocols. As one can see, the OC (red open circles) outperforms the reactive strategies  at all times,  showing that it is not just making the "good" moves when the final horizon is approaching.  \\
\textbf{Robustness.} To test the robustness of the different protocols,  we repeated a subset of successful catching episodes  with slightly different initial conditions. For the OC  strategies we kept the same history of the unperturbed steering protocols, while for the heuristic protocols  we allowed to change the reactive control according to the local environment along the perturbed trajectory (see Fig.~\ref{fig:sensitivity}a)  for a 3d rendering of a typical numerical experiment comparing OC and  PP strategies). 
Figure~\ref{fig:sensitivity}b displays the PDF of the final distances reached at catching time of the OC unperturbed reference episode  by the perturbed OC agent and by the agents following the three heuristic strategies. Results are shown for a subset of the hardest episodes where the unperturbed capture time is much larger than the mean, i.e. $t_c^{(\text{OC})}\gg 9\tau_\eta$, but still all the unperturbed heuristic trajectories capture within the time horizon. OC shows superior robustness against the reactive strategies concerning both the percentage of perturbed trajectories that succeed to capture the target and the final distance from the target for those episodes that do not succeed, as shown by the most pronounced peak  at the normalized capture distance, $R/\eta= 10^{-2}$ and by the much shorter right tail developed by the OC episodes with respect to the three reactive protocols (red circles).  
This result is particularly striking because the agents which use heuristic strategies start from perturbed positions but do follow the instantaneous "correct" control. The  performances of the three heuristic strategies would be even worst if we had used, as for the OC agents, the unperturbed  strategies learned along the unperturbed trajectory.  OC protocols appear to be quite robust, making the target's trajectory an attracting set for the controlled dynamics. This is further quantified in  Fig.~\ref{fig:sensitivity}{c} where we show for OC and PP strategies the  finite time Lyapunov exponent (FTLE)  evaluated as a function of the unperturbed capture time, $\gamma_{t^{(\text{OC})}_c}=1/t^{(\text{OC})}_c \ln(R_{t^{(\text{OC})}_c}/R_{0})$ where $R_{t^{(\text{OC})}_c}$ denotes the distance between the perturbed agent and the target \cite{brunton2010,krishna2022,krishna2023finite}. The FTLE is an increasing function of $t^{(\text{OC})}_c$, confirming that larger catching time are connected more chaotic  target trajectories. Moreover, FLTE for OC is always smaller than the one for  PP, independently of $t^{(\text{OC})}_c$, confirming that OC finds `more stable' trajectories. The findings presented here illustrate the impact of a perturbation in the initial condition of the order of $R_c$ (i.e., $\delta R_0 = R_c = \eta/100$). Anyway, considering other perturbations and limiting them to ensure the velocity field between the pair remains smooth, yield qualitatively similar results. 

\section{Discussion and Outlook}
We have shown how optimal control theory can be efficiently applied to control the dynamics of Lagrangian pairs in turbulent flows. Through the implementation of an iterative algorithm,  designed for solving the Euler-Lagrange equations arising from the Pontryagin Minimum Principle \cite{bryson,lenhart2007optimal,trelat2012optimal}, we have demonstrated how to catch a moving target in the shortest possible time or, at worst, how to limit the effects of chaotic dispersion. 

We focused on the relative dynamics of two Lagrangian objects in 3d turbulence, in the limit where their separation is always smaller than the flow dissipative scale. Both the moving target and the agent are carried by the turbulent flow, but the latter is equipped with some limited propulsion capabilities. Tracking a particle in turbulence is extremely challenging because of the exponential rate of separation due to chaos, making not obvious a priori that the iterative algorithm converges to an optimal solutions. Indeed, each point of an optimal trajectory depends on the entire history (past, present, and future) of the control and, consequently, a high fluctuation of the velocity field at a certain time may affect the entire trajectory, hindering the identification of optimal solutions. Notwithstanding these difficulties, we were able to obtain a high rate of convergence for the iterative algorithm, paving the way for the application of optimal control theory 
 -- so far mostly limited to simple, analytically solvable problems -- to turbulent setups as well.

 Converging to the optimal solutions requires perfect observability and full knowledge of the system dynamics. One may wonder about the trade-off between hard-to-get but far-sighted optimal solutions and easy-to-apply reactive protocols as, e.g.,  heuristic controls based only on local and instantaneous information of the environment. In this context, we developed a new reactive-control strategy by means of a perturbative approach to solve the optimal control problem. This strategy, which we dubbed \textit{perturbative optimal control}, aims to achieve the best possible performance when operating within the constraint of a short-sighted environmental evaluation.
 Owing to the assumption of persistent gradient it can be 
 handled analytically.
 Higher-order expansions of this perturbative approach can in principle be applied, leading to other new heuristic strategies. We  found that trading off observability with computational ease inevitably leads to a high loss in performance: more complex control strategies that just reactive ones appear to be necessary.   To fill the gap between optimal solutions and heuristic strategies, it would be interesting to improve the latter by introducing short-term memory with some analytical approximation or exploiting data-driven methods. 

 Remarkably, OC strategies showed to be highly resilient to disturbances on the agent starting conditions, which highlights how the drifting target trajectory becomes an attractor for the controlled dynamics. A future perspective would be to investigate how OC and heuristic strategies are impacted by errors in measurements (e.g. on position of the Lagrangian pair and on the local velocity gradients) that occur throughout the trajectory, as opposed to errors affecting only the initialization,  involving tools from stochastic optimal control theory \cite{fleming2012deterministicstochastic,CRESPO20032109}.  

 As a possible extension of the present results, with an eye to applications such as the control of micro-swimmers at small scales, it could be of interest to include hydrodynamic interactions between particles into the dynamics, as well as to take into account the impact of a non negligible reorientation time for the control variable.
 
 Finally, an outstanding issue is how to extend the search for moving targets to other situations of interest, for instance when separations are beyond the Kolmogorov scale and the velocity differences are not smooth. Even though heuristic, reactive, and local methods can in principle be employed, gauging their effectiveness is difficult in absence of the benchmark of the optimal control solution. In fact, how to apply optimal control theory in the inertial regime of turbulence remains an open problem, as iteratively solving of the Euler-Lagrange equations would require either the simulation of a new 3D Eulerian field at each iteration or the storage of the entire field's evolution. Both approaches appear to be computationally out of reach, hence the call to develop new numerical techniques to accurately solve large-scale Lagrangian optimization problems. 

\section{Methods}
\noindent\textbf{Navier-Stokes simulations for Lagrangian tracers}\\
 The target trajectories follow the tracer dynamics
\begin{equation}
    \dot{ \bold X}_t =  \bu_t(\Xtargett) \,,
\end{equation}
where $\bm u$ is a solution of the Navier-Stokes equations \cite{frisch1995turbulence,pope2000turbulent}
\begin{equation}
\label{eq:nse}
\begin{cases}
\partial_t \bm{u} + \bm{u} \cdot \nabla \bm{u}  = - \nabla p  
+ \nu\Delta\bm{u} + \mathbf{F} \\
\nabla \cdot \bm{u} = 0
\end{cases}
\,,
\end{equation}
for an incompressible fluid of viscosity $\nu$. The flow is
driven to a non-equilibrium statistically steady state by a homogeneous and isotropic forcing,  $\mathbf{F}$, obtained via a second-order
Ornstein-Uhlenbeck process \cite{forcingsawford}.
For the direct numerical simulations (DNS) we used a standard pseudo-spectral solver fully dealiased with the two-third rule. Details on the simulation can be found in \cite{TURB-Lagr,buzzicotti2016}. Parameters of the DNS used in this work are given in Table~\ref{tab:parameters}. The database of Lagrangian trajectories used in this study is dumped each $15 dt\simeq \tau_\eta/10$. 
\begin{table}[h!]
\centering
\begin{tabular}{|c|c|c|c|}
\hline
$N$  & $L$ &  $dt$  & $\nu$   \\  
1024 & $2\pi$  & $1.5\times 10^{-4}$  & $8 \times 10^{-4}$ \\ \hline
$\epsilon$    & $\tau_\eta$     & $\eta$ & $Re_\lambda$\\
 $1.4 \pm 0.1$ & $0.023\pm0.003$ & $0.0042 \pm 0.0001$ & $\simeq 130$ \\
 \hline
\end{tabular}
\caption{Parameters of the DNS: $N$ resolution in each dimension; $L$ physical dimension of the 3-periodic box; $dt$ time step in the DNS integration; $\nu$ kinematic viscosity; $\epsilon = \nu \langle \partial_i u_j \partial_i u_j \rangle$; $\tau_\eta = \sqrt{\nu/\epsilon}$; $\eta = (\nu^3/\epsilon)^{1/4}$; $Re_\lambda = u_{rms}\lambda/\nu$, where $\lambda = \sqrt{5E_{tot}/\Omega}$ is the `Taylor-scale' measured from the ratio between the mean system energy and enstrophy.}
\label{tab:parameters}
\end{table}

\noindent\textbf{Optimal Control equations}\\
Using the assumption that the distance between the agent and the
target is always within the scale of smoothness of the velocity field,
i.e. $R_t\lesssim 10\eta$, we can linearize Eq.~\eqref{eq:dynamics}
obtaining
\begin{equation}
\begin{cases}\label{eq:linear_dynamics}
    \dot{\bm R}_t=\nabla \bu_t \bm R_t\ + \bU_t, \\
    \bU_t=V_s\hat \bn_t\,.
\end{cases}
\end{equation}
Here, the pair separation $\bm R_t$ represents the state variable with a given initial
condition $\bm R_{0}$ and $\bU_t$ is the control variable. As
discussed in main text, we aim at solving the following optimization
problem: to find the best control that allows the argent to reach the
capture distance $R_c$ in minimal time $t_c\leq t_f$, where $t_f$ is a
fixed time horizon; if the capture is not realized, we require the
control to minimize the final distance, $R_{t_f}$, from the target.
The above twofold goal can be formally imposed by requiring that the control
minimize the following performance index
\begin{equation}\label{eq:performance_index}
    J=R^2_{t_f}+c\lambda R_c^2\,\int_{0}^{t_f}{dt\, f( R_t)}\,,
\end{equation}
where $\lambda$ is the uncontrolled Lyapunov exponent and $f$ is an
appropriate smooth function (see below) that is equal to $1$ for
$R_t>R_c$ and 0 for $R_t\leq R_c$. The first term in
\eqref{eq:performance_index} amounts to requiring minimal distance at
time $t_f$, while the second fullfils the request of minimal time to
reach the capture distance. The non-dimesional parameter $c$ weighs
the importance of the two objectives. Note that if the capture is not
reachable, the second term in the performance index is always a
constant, $c\lambda R_c^2 t_f$ and the problem remains minimize the
final distance. Conversely, if capture is realized the first term is
fixed to $R_c^2$. Therefore, to balance the two terms in such a way to
favour the capture we must choose $c> 1/(\lambda t_c)$, e.g. using
$t_c\sim 1-10 \tau_\eta$ we can estimate $c\geq 10-100$. The results shown here correspond to $c=100$; while higher values of $c$ lead to same results, $c\sim O(1)$ or even smaller does not allow to find optimal strategies.
Given the unconstrained performance index
\eqref{eq:performance_index}, by imposing the dynamics
\eqref{eq:linear_dynamics} and the non-linear constrain in the control
variable, $|\hat \bn_t|^2=1\, \forall t$, we are left with  the following
constrained optimization problem
\begin{eqnarray}
\label{eq:oc}
\tilde J &=& R_{t_f}^2 + \int_{0}^{t_f} dt\, \left[c\lambda R_c^2\,f(R_t) \right.+ \\
&+& \left. \bm \phi_t\cdot(\nabla \bm u_t \bm R_t+ V_s\hat{\bm n}_t-\dot {\bm R}) + \mu_t(1-|\hat{\bm n}_t|^2)\right]\,,\nonumber
\end{eqnarray}
where  $\bphi_t\,,\mu_t$ are the Lagrangian multipliers with the role of co-state \cite{bryson}. Integrating by parts, and requiring the stationarity of $\tilde{J}$ upon variation  of control $\hat{\bm n}_t$ and state $\bm R_t$ the optimization reduces to solving the following "Euler-Lagrange" equations \cite{bryson}: 
\begin{eqnarray}
\dot{\bm \phi}&=&-\tfrac{\partial H}{\partial \bm R}\,,\label{eq:EulerLagrangeA}\\
\tfrac{\partial H}{\partial \hat \bn}&=&0\,,\label{eq:EulerLagrangeB}
\end{eqnarray}
$H=c\lambda R_c^2 \,f(R_t)+\bphi_t\cdot(\nabla\bu_t\bm R_t + V_s \hat{\bm n}_t)+\mu_t(1-\hat \bn_t^2))$ being the Hamiltonian function of the constrained minimization problem. The equation for $\dot{\bm \phi}$ has final condition $\bm \phi_{t_f}=2\bm R_{t_f}$, while the dynamics of the state variable \eqref{eq:linear_dynamics} (which, as discussed below, is actually modified for stopping the dynamics when the capture distance is reached) has initial condition $\bm R_0$. Notice that Eq.~\eqref{eq:EulerLagrangeB} prescribes the control to be
\begin{equation}
  \label{eq:controlupdate}
   \hat \bn_t = \frac{V_s\bphi_t}{2\mu_t}=-\frac{\bphi_t}{|\bphi_t|}\,,
\end{equation}
where $\mu_t$ plays the role of normalization factor and where the
minus sign is to impose $\frac{\partial^2 H}{\partial \hat \bn^2}>0$
as we are performing a minimization.

In practice, to find the optimal control solution that minimizes
$\tilde{J}$, we use the Forward-Backward Sweep
Method (FBSM) \cite{lenhart2007optimal}.
At first, it requires an initial guess for the control variable $\hat \bn_t$ for $ 0\leq t\leq t_f$. Then, the problem becomes
computationally heavy since it requires iterative searching with
backward (for the Lagrangian multipliers) and forward (for the state variable)
integration such as to identify the optimal control. The algorithm can be
summmarized with the following pseudocode 

\begin{table}[H]
\begin{tabular}{l} 
\hline
\textbf{Algorithm:} Forward-Backward Sweep  \\
\hline
Parameters: learning rate $\gamma$, threshold for convergence $\delta$\\
\quad$1:$ Initial guess for the control variable $\hat{\bn}_t\, \forall t \in [0,t_f]$\\
\quad$2:$ \textit{Forward} integration of $\bm R_t$ in $t\in [0:t_f]$ (Eq.~\eqref{eq:linear_dynamics})\\
\quad$3:$ \textit{Backward} integration of $\bm \phi_t$ in $t\in [t_f:0]$ (Eq.~\eqref{eq:EulerLagrangeA})\\
\quad$4:$ \textit{Update control} $\hat{\bn}_t \leftarrow (1-\gamma)\hat{\bn}_t+\gamma \frac{\bphi_t}{|\bphi_t|}$ (Eq.~\eqref{eq:controlupdate})\\
\quad$5:$ \textit{Check convergence} \textbf{if}  $\sum_{i=1}^{t_f/dt}|\Delta w_{i\,dt}| \leq \delta \sum_{i=0}^{t_f/dt}|w_{i\,dt}|$ \\
\quad\quad\; for all $\bm w\in\{\bm \phi,\bm R, \hat{\bm n}\}$ end \textbf{else} goto 2\\
{\tiny where $\Delta w_t$ is the difference between the old and new estimate of the variables}\\
\hline
\end{tabular}
\end{table}

In the practical implementation we used $\gamma=5\times 10^{-4}$, $\delta=10^{-4}$ and as initial guess for the control variable we used the best heuristic strategy, i.e., the one with the minimum performance index. Using PP for the initialization provides quantitatively similar results.
If the convergence is not realized after 20K iterations of the algorithm, we repeated the FBSM rescaling the parameters $\gamma\leftarrow \gamma/10, \delta\leftarrow \delta/10$ and increasing the maximum number of iteration to $200$K.  This setup ensure the convergence in the $80\%$ of the total episodes studied. In the other $20\%$ of episodes, the FBSM algorithm does not converge, so the global optimum is not achieved. However, it is always possible to consider the strategy that during the iterations of the FBSM algorithm provides the minimum of the performance index as an approximate solution of the optimal control.  Indeed, since the initialization of the algorithm is provided by the best of the heuristic strategies, the approximate solution will always be an upper bound of the reactive behaviour. By refining the hyper-parameters even more, should be possible to enhance the algorithm's convergence rate to potentially reach $100\%$.  Moreover for the function $f$ in Eq.~\eqref{eq:performance_index} we used 
\begin{equation}
    f=\tfrac{1}{2}\tanh\left(\alpha\tfrac{R_t-R_c}{R_c}\right)
\end{equation}
which is a smoothed version of the Heaviside function, whose stiffness
is ruled by $\alpha$ (here chosen to be 10). In addition, to impose the capture condition, i.e. that whenever the agent is at distance $R\leq R_c$ from
the target it sticks to it we modified Eq.~\eqref{eq:linear_dynamics} such that $\dot{\bm R} \to 0$ when $R\leq R_c$ by redefining the dynamics as follows:
\begin{equation}
    f\,\dot{\bm R} \to \dot{\bm R}\,,
\end{equation}
which implies that the backward evolution of the Lagrangian multipliers in \eqref{eq:EulerLagrangeA} becomes
\begin{equation}
    \dot \bphi = -\Big{[}c\lambda R_c^2 + \Big{(}\nabla \bu_t \bm R_t + V_s \hat \bn_t\Big{)}\cdot \bphi_t\Big{]}\frac{\partial f}{\partial \bm R} - f {\nabla {\bu}^\intercal_t} \bphi_t\,. 
\end{equation}
The rate of convergence of the algorithm is still close to the $80\%$ by increasing/decreasing the initial conditions and changing the velocity amplitude accordingly as, e.g., considering $(10 R_0\to R_0,  10V_s\to V_s)$ or $(R_0/10\to R_0, V_s/10\to V_s)$. 
\newline\newline

\noindent\textbf{Heuristic Strategies}\\
Differently from the optimal control equations, the heuristic control strategies we studied are reactive, meaning  that they need only instantaneous information about the system to be applied. The pure pursuit (PP) strategy, for instance,  does not exploit any information on the flow during the navigation but constantly realign the control direction to the moving target, i.e. $\hat{\bm n}^{\text{PP}}_t = - \hat {\bm R}_t\,$.  
The other two strategies, the surfing control (SC) and the perturbative optimal control (PO), instead, consider both the direction of the target and the instantaneous velocity gradient. While these two strategies are both based on a free parameter to be optimized numerically, $\tau_s$ and $\tau_p$ respectively, they are obtained from different assumptions. \\
\textbf{Surfing control strategy}\\
This control is inspired by Ref.~\cite{monthiller2022surfing}, where one assume that a linear approximation of the flow underlying the active particle is reasonable for an interval of time $\tau_s$.  In other words, one has to assume $\tau_s$ as the persistence time of the local (to the agent) gradients of the flow. This control was proposed as an effective strategy to drift in a given direction (e.g. the vertical one) fixed in time, essentially it is obtained with the request to maximize over the time interval $\tau_s$ the displacement in the chosen direction.
Here, we adapt it to our case by assuming that the direction toward the target, $\hat{\bm R}_t$, remains  constant during the same interval of time $\tau_s$. Then, maximizing the searcher displacement along the target direction, i.e. $\max_{\hat{\bm n}_t}\big{[}-(\bold X^{(a)}_{t+\tau_s}-\bold X^{(a)}_{t})\cdot \hat{\bm R}_{t}\big{]}$ and considering a continuous measurement of the environment, we find the following  strategy:
 \begin{equation}
    \hat {\bm n}^{\text{SC}}_t = -\frac{[\exp(\tau_s \nabla \bm u_{t})]^\intercal \hat{\bm R}_{t}}{| [\exp(\tau_s \nabla \bm u_{t})]^\intercal \hat{\bm R}_{t}|}\,.
\end{equation}
The details on the derivation are discussed in Ref.~\cite{monthiller2022surfing} and are not repeated here.\\

\noindent\textbf{Perturbative optimal control strategy}\\
The novel  PO  strategy we propose, works in the same regime we derived optimal control equations, considering the velocity field between the agent and the target to be linearizable \eqref{eq:linear_dynamics} but, similarly to the surfing, by assuming local gradients persistence for a time $\tau_p$. Under these assumptions, it is easy to show that  
\begin{equation}
    \bm R_{\tau_p}= \exp(\tau_p\nabla \bu_{0})\bm R_{0}+ V_s\int_{0}^{\tau_p}{dt\,\Big{[}\exp[(\tau_p-t)\nabla \bu_{0}]}\,\hat \bn^{\text{PO}}_t \Big{]}\,.
\end{equation}
Then, we derive the PO strategy by imposing that $\bm R_{\tau_p}\cdot \bm R^{V_s=0}_{\tau_p}$ is minimum (note that this perturbative approach can be thought as the $0^{th}$ order, in $V_s$, solution of the optimal control equations), i.e., minimizing directly the separation of the pair. Finally, assuming continuous measurements as before, we obtain the following control:
\begin{equation}
    \hat{\bm n}^{\text{PO}}_t=-\frac{[\exp(\tau_p{\nabla \bu}_{t})]^\intercal\exp(\tau_p\nabla \bu_{t} )\hat{\bm R}_{t}}{| [\exp(\tau_p{\nabla \bu}_{t})]^\intercal\exp(\tau_p\nabla \bu_{t} )\hat{\bm R}_{t} |} \,.
\end{equation}
\\

Note that SC and PO strategies recover the PP for the free parameters, $\tau_s$ and $\tau_p$, set to be zero. 
Clearly, the optimal free parameters depend on the temporal variation of both the underlying gradients, $\nabla \bm u_t$, and the target direction, $\hat {\bm R_t}$. Their values have been determined empirically; we have searched for the values that maximize the capture frequencies of $2\times10^5$ different Lagrangian target, when $\tau_s$ and $\tau_p$ varied in the range [$0,4\tau_\eta$]. In particular, we have found as optimal parameters $\tau_s\simeq 0.6 \tau_\eta$ and $\tau_p \simeq 1.3\tau_\eta$. 

\section{Data and Code Availability}
The Lagrangian target trajectories (including positions, velocities, accelerations and fluid gradients along each particle) used in this work are available for download in the Smart-TURB portal \url{http://smart-turb.roma2.infn.it}, under the TURB-Lagr repository \cite{TURB-Lagr}. TURB-Lagr is a new open database of 3d turbulent Lagrangian trajectories, obtained by Direct Numerical Simulations (DNS) of the Navier-Stokes equations with  homogeneous and isotropic forcing.  Details on how to download and read the database are also given in the portal. 
The analysis that support the findings of this study are available from the corresponding author upon reasonable request.

The code (written in C language) to study the optimal tracking strategies in a turbulent flow is free downloadable on GitHub at this link: \url{https://github.com/calascibetta-chiara/Optimal-tracking-strategies-in-a-turbulent-flow}
\newline

\textbf{Correspondence} and requests for materials should be addressed to Chiara Calascibetta.

\section{Acknowledgements}
This work was supported by the European Research Council (ERC) under the European Union’s Horizon 2020 research and innovation programme (Grant Agreement No. 882340).
\section{Author contributions}
All authors conceived the research. C.C. performed all the numerical simulations and data analysis. All authors contributed to the interpretation of the results and writing of the manuscript.   

\section{Competing interests}
The authors declare no competing interests.

\bibliographystyle{naturemag}

\end{document}